\documentclass[11pt]{article}
\usepackage[top=1in, bottom=1in, left=1in, right=1in]{geometry}

\AtBeginDocument{%
  \providecommand\BibTeX{{%
    \normalfont B\kern-0.5em{\scshape i\kern-0.25em b}\kern-0.8em\TeX}}}

\usepackage{tcolorbox}
\usepackage{amsmath}
\usepackage{parskip}
\usepackage{booktabs}
\usepackage{wrapfig}
\usepackage{enumitem}
\usepackage{natbib}
\usepackage[hyphens]{url}
\usepackage{hyperref}

\newcommand{\anon}[1]{ANONYMIZED} 
\newcommand{\Switzerland}[0]{{\texttt{Switzerland}}}
\newcommand{\Valais}[0]{{\texttt{Valais}}}
\newcommand{\AC}[0]{{\texttt{Appel Citoyen}}}
\newcommand{\Monthey}[0]{{\texttt{Monthey}}}

\newcommand{\ANONURL}[0]{{\url{https://github.com/huanglx12/Balanced-Committee-Election}}} 

\newcommand{\ccolor}[1]{\textcolor{black}{#1}}

\usepackage{authblk}

\title{Diverse Representation via Computational Participatory Elections - Lessons from a Case Study}

\author{{Florian Evéquoz}}
\affil{University of Applied Sciences of Western Switzerland HES-SO}
\affil{{Human-IST Institute, University of Fribourg}}
 
\author{{Johan Rochel}}
\affil{University of Zurich}

\author{{Vijay Keswani}}
\author{{L. Elisa Celis}}
\affil{Yale University}

\date{}

\begin{document}




\maketitle

\begin{abstract}
  Elections are the central institution of democratic 
  processes, and often the elected body -- in either public or private governance -- is a committee of individuals.
  To ensure the legitimacy of elected bodies, the electoral processes should guarantee that diverse groups are represented, in particular members of groups that are marginalized due to gender, ethnicity, or other socially salient attributes.
  To address this challenge of representation, we have designed a novel participatory electoral process coined the \textit{Representation Pact}, implemented with the support of a computational system. That process explicitly enables voters to flexibly decide on representation criteria in a first round, and then lets them vote for candidates in a second round. After the two rounds, a counting method is applied, which selects the committee of candidates that maximizes the number of votes received in the second round, conditioned on satisfying the criteria provided in the first round. %
  \ccolor{
  With the help of a detailed use case that applied this process
  in a primary election of 96 representatives in \Switzerland{}, we explain how this method contributes to fairness in political elections by achieving a better ``descriptive representation''. 
  Further, based on this use case, we identify lessons learnt that are applicable to participatory computational systems used in societal or political contexts.}
  Good practices are identified and presented.
\end{abstract}



%

%
\section{Introduction}

Election is a high point of democracy. Citizens select the people who will represent them and enact laws on their behalf. However, marginalized groups in a population are rarely well-represented among the elected. %
%
This gap between a population and its elected representatives undermines the trust that citizens have in the decisions that are made on their behalf~\citep{mansbridge_what_2000,arnesen_legitimacy_2018}. 

To address this challenge of representation, we have designed a novel participatory electoral process coined the \textit{Representation Pact}, implemented with the support of a computational system. This process involves two sequential phases. In a first phase, voters collectively define certain goals of the election, i.e. by setting what representation criteria should be respected in the election result (e.g by reserving seats for specific gender, age or ethnicity). These criteria are then integrated in the design and configuration of the computational system involved in the counting of ballots. In a second phase, voters choose their preferred candidates. After the two phases, the computational system configured with the choice of voters is used to identify winners, taking into account the representation criteria and the votes. The Representation Pact improves democratic legitimacy as the representation criteria  are determined by the voters themselves in a first phase, before being applied to the election phase.  Overall, we argue that the Representation Pact contributes to a fairer electoral system, thereby improving the legitimacy of elected bodies and the quality of democratic decision-making procedures.

Our contribution is sociotechnical and applied in nature. It sits as the junction of political philosophy, computer sciences, and algorithmic law scholarship and discrimination. 
We apply here a computational system in the context of an electoral process. {Along similar lines as the} scholarship addressing discriminatory risks in the conception and impact of AI systems, we consider a computational system integrated into a political mechanism as a way to promote more balanced elected bodies. Our work is an example of developing a computational system for a specific political application that advances important normative goal (representation). As we shall argue, the fact that this system can be  configured in a participatory manner by the choice of voters prevents challenges related to ``algocracy''~\citep{danaher_threat_2016}.

To fully make sense of this proposition, we firstly need to address the role of representation in democracy from a political philosophy perspective, and argue that the Representation Pact offers a practical instrument for ensuring ``descriptive representation'' with strong democratic guarantees. {We  discuss its theoretical benefits and limitations from a social justice perspective, as well as, the motivations and implementation challenges from a participatory design perspective. }
Secondly, we present a case study that translates theoretical concepts and algorithmic models into political practice. We describe the process of applying the computational system to a real election: a political movement’s primaries that allowed 1,905 registered voters to elect 96 representatives (among 151 candidates). 
In addition to the example usage of \ccolor{the participatory computational system} in electoral processes, this case study offers an opportunity to learn about criticisms brought by citizens facing the use of such computational system in elections. We discuss them at the end of the article, before providing guidelines and recommendations for the application of the Representation Pact.

{
Our work extends the ever-growing literature on employing technical tools to incorporate diversity in democratically-elected committees.
%
Even beyond research, multiple countries have already adopted procedures that constrain the outcomes of multi-winner elections based on pre-defined representation rules.
This includes implicit mechanisms to guarantee representation by selecting political party/ideology-diverse committees, for instance, via proportional representation rules employed in national and local council elections in Switzerland \cite{lutz2004switzerland}, and national elections in New Zealand \cite{nagel2000expanding}.
Or explicit methods such as reservation and quota systems in India \cite{jensenius2013power} and Brazil \cite{jones2009gender}.
Both kinds of methods aim to induce diversity in the selected committees and are often assisted by computational tools.
%
%
%
%
%
%
%
However, existing implicit mechanisms, like proportional representation rules, do not always lead to committees where marginalized communities are well-represented 
\cite{ijcai2018-20,bredereck2018multiwinner}
, often necessitating the use of explicit methods to guarantee diversity.
Our work addresses the questions around defining explicit diversity rules and procedures to incorporate these rules in the election process.
%
\ccolor{Further, through our presented case study, we provide details of the technical and practical challenges faced in a real-world implementation of our framework, complementing the prior work in this field which primarily show efficacy of their approaches in synthetically-simulated settings \cite{ijcai2018-20,skowron2016finding,lu2011budgeted,lang2018multi}.
}
%
We propose a collaborative process of election design that integrates different stakeholders' views on the desired diversity in elected bodies and provides a feasible mechanism to incorporate this diversity in the voting process.
}



\section{Related Work}
\label{sec:relatedwork}
Our contribution fits nicely into the current literature in political philosophy and political science and connects these fields with the computer science 
literature. 
In political philosophy, our work can be seen in the continuation of the seminal work by~\citet{young_justice_1990}. Our contribution might be read as a specific example of institutional design research, such as that promoted by consociationalism~\citep{andeweg_consociationalism_2016,lijphart_democracy_1977,lijphart_evolution_2002}. As proposed by~\citet{stojanovic_dialogue_2013}, this design research should be linked to federalist research. This again brings us back to more philosophical issues related to the issue of fairness in multilingual/multinational states~\citep{kymlicka_multicultural_1995,van_parijs_linguistic_2011}.

These conceptual discussions both enrich and are enriched by the mathematical and theoretical computer science literature. Scholars from these domains have clarified in particular how to address the proportion of representation within a criterion. In particular, the types of criterion we discuss could encode notions of fairness such as statistical parity~\citep{dwork2012fairness} or other notions of diversity such as the 80\% rule~\citep{DBLP:conf/fie/CohoonCRL13,reyland2017how}.
Additionally, the criterion could also be set to generalize notions of proportionality that have arisen in the voting literature such as fully proportional representation~\citep{monroe1995fully}, fixed degressive proportionality~\citep{koriyama2013optimal}, and flexible proportionality~\citep{brill2017multiwinner}.
\ccolor{Recent work has also studied the impact of these notions in real-world settings; \citet{cembrano2021proportional} assess the impact of different kinds of synthetically-generated proportional representation rules on representativeness, robustness, and voting power of the chosen committee using the data from 2021 Chilean
Constitutional Convention election.
In contrast to this paper, our framework studies a real-world setting where the voters select the representation criteria. 
}
%
Importantly, the fairness criterion also allows representation across multiple groups by adding multiple criteria -- this is fundamentally new in the voting literature; the above examples and solutions can only handle a single criterion. Thanks to this progress, we can advance a real argument about the different components of a fair political election. With a single criterion, the normative discussion was too poor. 
\ccolor{With the capability of handling multiple criteria, there is a true opportunity to link normative arguments about fairness and potential implementations in the form of mathematically secured processes.
}

The suggested type of criteria solutions also has connections to state-of-the-art research in attaining fairness in algorithmic systems. 
Constraints (which translate to criteria)  have been studied by a number of works on fairness generally~\citep{celis2019classification},
including in several works that aim to rank which criteria have direct parallels to the voting case if we consider the vote 2 as a ranking of candidates~\citep{celis2018ranking, linkedin_ranking_paper,ZehlikeB0HMB17}.
In~\citet{ijcai2018-20} it is shown that a slight amount of flexibility in the constraints (e.g., at least 40\% male winners and at least 40\% female winners, as opposed to a strict 50\%-50\%) can significantly improve the score. 
{
Other works in the field of computational social choice \cite{meskanen2008closeness,lang2018multi,faliszewski2017multiwinner} and algorithmic fairness \cite{bredereck2018multiwinner} have similarly proposed technical frameworks for finding solutions to constrained multi-winner problems.
}

Finally, user and stakeholder participation in the design of algorithmic systems has received a growing attention in recent years~\citep{wolf_changing_2018}. \citet{lee_webuildai_2019} proposes a framework that enables people to participate in the collective designing of algorithmic policies for their communities. Our work shares the same purpose of letting people participate in the definition of algorithmic properties and constraints, although in a simpler, less generic, process that requires only deliberation and voting to adopt representation criteria in elections.
{
Other general participatory design principles have been used in various civil contexts to incorporate the advantages of increased community participation.
{
\citet{pilemalm2018participatory} discuss the use and applicability of participatory design principles towards involving citizens and broad masses in public service organizations, with the overall goal of increasing efficiency and effectiveness of services provided by these organizations.
\citet{moore2016participatory} similarly suggest that citizen participation can address the problems associated with lack of data in urban planning projects.
Various other tools have also been proposed to to facilitate data collection and sharing by local and global communities for collaborative initiatives \cite{lee_webuildai_2019,balestrini2017city,matias2018civilservant,vlachokyriakos2014postervote}.
Our work similarly provides a collaborative framework to ensure representation in the outcomes of multi-winner elections.
}
A detailed discussion of the participatory design and philosophical grounding of our work is presented in Appendix~\ref{sec:motivation}.
%
}

\section{The Representation Pact}
\label{sec:pact}
Election processes are designed to ensure representation across certain criteria. For instance, they can guarantee representation in relation to such criteria as (1) geographical criterion (fixed number of representatives per geographical unit) and (2) partisan membership criterion (proportional representation of political forces according to the votes obtained). In the United States of America, at the federal level there is geographical representation in two forms: by county (representing roughly a fixed number of people) in the House of Representatives, and by state (representing a varying number of people) in the Senate. However, there is no explicit representation criteria according to political party affiliation in either.
In the Republic of India, on the other hand, specific seats are reserved in the House of the People for members of certain Scheduled Tribes and Scheduled Castes which have faced historic discrimination. These representation rules are often embodied in the Constitution of the countries. They provide stability, but are unlikely to change dynamically or be extended in order to cope with the evolution of society and new expectations in terms of representations (e.g. according to gender, ethnicity, minorities, etc.) 
The objective of our work is thus to define an electoral process based on rules that are binding, modifiable and legitimate. Citizens must be able to discuss the criteria of representation, make an explicit choice about them, and fully agree to apply the chosen criteria during elections. We call this approach the Representation Pact.

\begin{figure}
  \centering
  \includegraphics[width=0.9\textwidth]{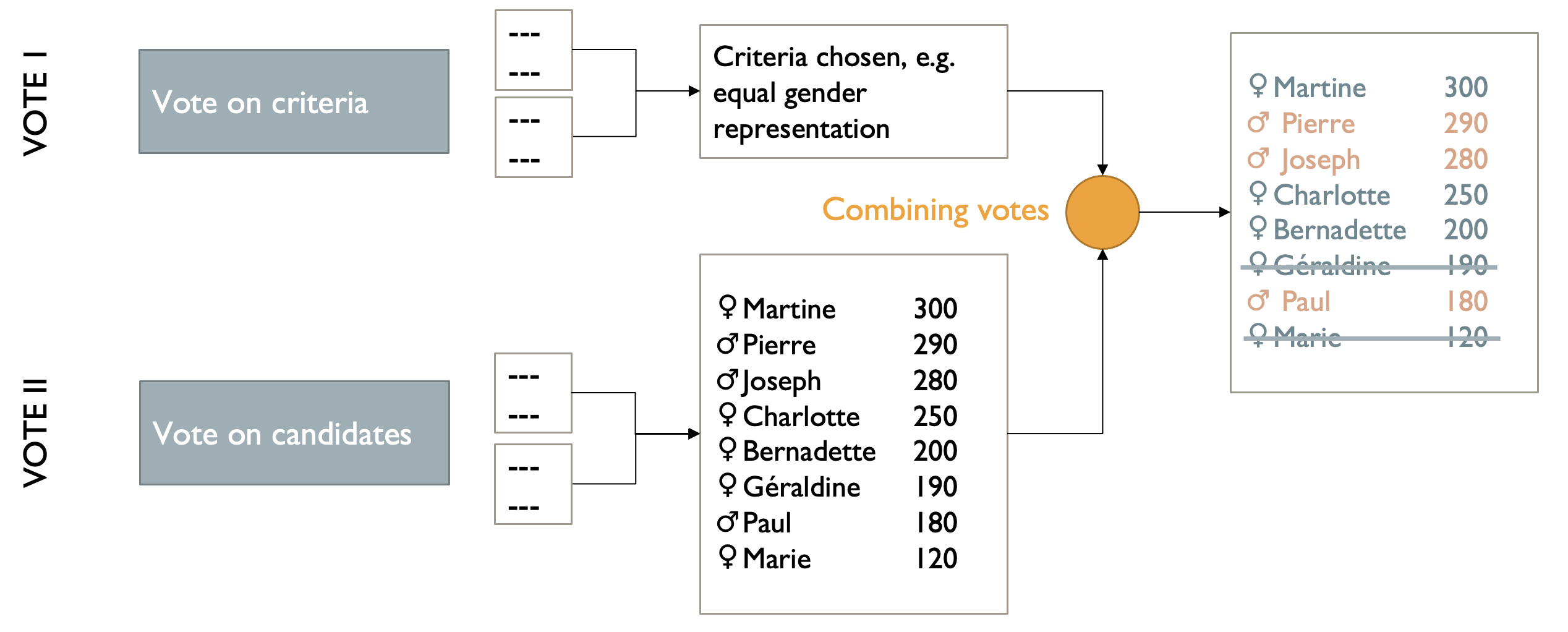}
  \caption{The Representation Pact consists of two separate votes. The first one allows voters to choose representation criteria. The second one lets them choose their preferred candidates. The counting method ensures that the agreed upon criteria are taken into account when choosing the winning committee while at the same time maximizing the total number of votes.
  }
  \label{fig:representation-pact}
\end{figure}

The Representation Pact is composed of two parts. 
The ``Pact'' reflects the need for citizens to debate and choose certain criteria. This choice takes the form of a binding commitment. The concept of \textit{representation criteria} refers to the specific categories chosen by citizens. In public discourse, these criteria are often referred to as  \textit{quotas}. 
``Representation'' refers to the ambition of determining collectively an ideal or desired representative outcome of an election. It is part of a broader conversation on fair political institutions, especially from the perspective of integrating political minorities. %
Together, these two votes form a complete democratic moment. Fig.~\ref{fig:representation-pact} illustrates the two phases of the Representation Pact, which we describe in detail below.
The Representation Pact could apply either to official elections, primary elections within a party, or to elections within a company or association. It is important to note that the Representation Pact does not require electronic voting. The approach is compatible with digital or analog voting methods. Only the result of the votes must be translated in a numerical format as an input to the counting algorithm.
%


\subsection{Voting on Criteria}

The first phase of the Representation Pact is used to determine the criteria of representation that the group commits to (the ``Pact''). It is a participatory approach to collectively choose the representation goals of the election, and effectively configure the underlying computational system.
The vote is formulated as a series of yes/no questions on the appropriateness of meeting a criterion, for example: 
\begin{itemize}[noitemsep,topsep=1pt]
    \item ``Would you like the list of party X for the next elections to contain equal numbers of male and female candidates?'' (primaries) 
    \item ``Do you want the Council to have the same number of male and female members?'' (official election). 
\end{itemize}
A positive answer to one or more of these questions leads to the application of the corresponding criteria in the second vote (the election). 
This first moment allows citizens to clarify which representation they wish to see achieved. As explained above, they determine which features of the representatives
are crucial for them to secure the quality of political decision-making. 
It is not a question of disrupting the entire political system, but of asking an essential question for a specific election: in terms of representation, what kind of candidates for the list/parliament/council do you want?$\\$
\ccolor{
The choice and formulation of the criteria poses many challenges. 
For instance, how should criteria be chosen, i.e., what attributes are to be used in the criteria and what are target representation values for these attributes? 
%
%
Answering these questions requires thoughtful consideration of the specific electoral process in question.
%
Through the case study in Section~\ref{sec:usecase}, we provide a concrete example of an appropriate representation criteria used in a real-world implementation of Representation Pact.
%
Nevertheless, the construction of the criteria can be highly context-specific, and we provide guidelines and recommendations to handle the complexity of this construction process in Section~\ref{sec:guidelines}.
}

\subsection{Voting on Candidates}

The second phase of the Representation Pact looks more similar to a standard election. 
Voters select the candidates they would like to see in office. 
This election might take the form of a majority vote with several candidates (plurality-at-large voting). 
The counting process gathers the individual votes received by all candidates. 
On this basis, a list of raw results is established (the number of votes received by each candidate individually). 

In the most usual form, if $k$ candidates are to be elected, it is the $k$ with the most votes that win.
The challenge is to ensure that the representation criteria chosen in the first vote are respected.
This must combine the choice of criteria expressed by the electorate (vote 1) and the election (vote 2). 
Simply taking the top $k$ candidates may violate the criteria from vote 1, and hence a new approach is required.

\subsection{Combining Criteria and Candidate Votes}

%
The goal is to determine a winning list such that, among all possible combinations of candidates who meet the criteria (vote 1), it has the most total votes received by the winning candidates (vote 2). 
In effect,  the choice of the winning list should not be arbitrary. The winning list must result from a mathematical and deterministic application of the criteria chosen upon the list of raw results. The organizers of the vote (the State for an official election, or a party for a primary) must not intervene and choose between several possible ``winning'' lists that satisfy vote 1. 
The system must therefore be designed to produce a single winning list: the group of candidates who meet the criteria and who, together, have more votes than all other possible groups of candidates who meet the criteria. 
In practical terms, this electoral innovation is made possible by a mathematical way of posing the problem of election (and solving it).

\ccolor{
The method employed to count the votes and find the winners is based on the Integer Linear Programming (ILP) paradigm.
Formally, we have $m$ candidates $C=\left\{1,\ldots,m\right\}$, a total of $n$ voters and a desired size $k\in \left\{1,\ldots,m\right\}$ of a committee to be elected. 
For the purposes of illustration, let's say each voter can select at most $k$ candidates, though this restriction can be changed without consequence. 
Let's say candidate $i$ receives a total number of votes $w_i$; we can compute the votes received by any committee $S\subseteq C$ of size $k$
by summing up all of the votes for the candidates in $S$, i.e. $\sum_{i\in S} w_i$.
%
%
Overall, our goal is to elect a committee $S\subseteq C$ of size $k$ that satisfies the criteria selected
%
and that has the most votes, i.e. maximal $\sum_{i\in S} w_i$, amongst all feasible committees.
This ILP with representation constraints can be solved in practice using standard publicly-available solvers, like \textbf{CPLEX} or \textbf{CVXPY}.
An implementation of our framework is publicly available at \ANONURL{}
%
%
%
and we provide additional details of our proposed ILP program in Appendix~\ref{sec:ilp}.
}

By focusing on the plurality-at-large case, a simple solution
can be derived once the specific setting is framed as an ILP; publishing the ILP and using a publicly available solver is crucial for transparency.
Modern solvers are sufficiently efficient in handling problem sizes encountered in the election described in the use case in Section~\ref{sec:usecase}. 
Thus, this approach is robust, accurate, and fast enough to be used in determining the outcome of a Representation Pact election.
\ccolor{By using a voter-selected criteria, the Representation Pact aims to achieve ``descriptive representation'', i.e. elected representatives are similar (in significant ways) to the people who have elected them. 
The philosophical motivations for achieving this form of representation through the Pact are discussed further in Appendix~\ref{sec:motivation}. 
}

\section{Use Case}
\label{sec:usecase}

{\Switzerland{} has a long established democratic tradition. People are usually called to vote several times a year on referendums and initiatives. The national parliament is elected every 4 years. Regional and municipal parliaments and governments are elected every 4 or 5 years depending on the State's laws. Elections of parliaments are held using a proportional representation scheme.}

The Representation Pact was used 
in \Valais{}, \Switzerland{} in the context of the election of a Constitutional Assembly, {to determine lists of candidates of a citizen movement}. This use case offers insights pertaining to computational explainability and transparency challenges in the context of a real political process. 
{The election of the Constitutional Assembly follows a proportional representation scheme, which requires candidates to be on party lists.} 
A number of citizens united to form an independent political movement called \AC{}
\footnote{Two authors of this paper were members of the board of \AC{}.}.  
96 seats in the Constitutional Assembly would be allocated to the 8 districts in which \AC{} had candidates. 
\AC{} would present one electoral list in each district. In order to open the possibility for anybody interested to become a candidate, \AC{} decided to organize primary elections (``primaries'') in each of those 8 districts to choose the people who would appear on their list. The primaries were organized according to the Representation Pact. 

After an internal consultation, the board of \AC{} proposed that three representation criteria, based on the demographics of the region, be put to vote:
\begin{itemize}[noitemsep,topsep=1pt]
\item Gender: 50\% male and 50\% female candidates overall. A difference of at most 1 in the number of male and female candidates in each district (odd number of candidates).\footnote{There were no candidates who did not identify as either male or female.}
\item Age: at least 10\% of candidates aged between 18 and 30, at least 10\% of candidates aged above 65, at least 40\% of candidates between 31 and 65 in each district.
\item Region: reserved seats  for municipalities in each electoral district, based on their population.
\end{itemize}

\noindent
Before the vote, the consequence of the choice of each representation criteria was made public. 
An example of that documentation for the district of \Monthey{} is shown in Table~\ref{tab:criteria} and the ILP for this district is presented in Fig.~\ref{fig:monthey} in the Appendix.
The vote was open to everybody, including foreigners living in \Valais{}. Interested people could register on an online platform, providing their name, email address, electoral district, along with a copy of their ID to allow for verification and prevent fraud. For that first vote, 453 people registered and 347 of them (76.6\%) participated in the vote that took place on a digital e-voting platform.
Questions were formulated as closed (yes-no-blank) questions. All three criteria were accepted, with the following scores:

\begin{itemize}[noitemsep,topsep=1pt]
    \item Gender: 74.9\% yes, 19.6\% no, 5.5\% blank
    \item Age: 76.9\% yes, 17.6\% no, 5.5\% blank
    \item Region: 70.0\% yes, 21.6\% no, 8.4\% blank
\end{itemize}

Then, citizens were invited to become candidates on \AC{} lists.
%
There was a clear and consistent communication that the three chosen criteria would be applied to the choice of the candidates. Any interested person 
could register at \AC{}, by providing their name, age, gender and place of residence. A list of candidates for each district, along with their characteristics, was made public before the election. 

A total of 151 candidates (77 male, 74 female) registered for the second phase of the election. Additionally, 1,905 registered and verified voters were allowed to vote in the second phase. The election took place three months after the first phase 
and 1,308 voters participated (68.8\%). Voters could nominate the candidates they preferred in each district. The election followed a plurality-at-large scheme. Voters in each district could vote for up to $S$ candidates, with $S$ being the number of seats allocated for that particular district in the general election. There were no other constraints on the vote. In particular, voters did not need to apply the chosen representation criteria to their individual ballot. These representation criteria were incorporated in the counting phase. 

At the end of the procedure, 96 candidates were nominated. In accordance with the objectives of the Representation Pact, the 96 candidates met the chosen criteria and thus reflected the diversity of the inhabitants of the canton:

\begin{itemize}
    \item Gender: 48 men, 48 women (50\% - 50\%)
    \item Age: 27 people aged 18-30, 54 people aged 31-65, 15 people aged more than 65 (28.1\%, 56.3\%, 15.6\%)
    \item Region: 40 municipalities represented by the candidates, out of 63 municipalities in the 8 districts where \AC{} had candidates (63.4\%)
\end{itemize}

To ensure transparency and trust in the process, raw anonymized ballots were published in an open format (CSV). Voters received a unique code or 'hash' known only to them after they cast their vote. This 'hash' was published alongside the ballots, allowing them to verify that their vote was taken into account. Moreover, the source code of the counting algorithm was published in order to allow anyone to reproduce the counting and verify it.

In order to exemplify the kind of effects that can appear in a real-world situation, let's take a closer look at the results of the election in the district of \Monthey{}. This district is especially interesting as it illustrates some of the tensions which are raised by the application of representation criteria, and challenges pertaining to 
explainability issues.

\begin{table}[]
    \centering
  \caption{Use case --- (Left) Precise categories and values of representation criteria in the electoral district of \Monthey{}, which has a total of 17 seats. This information was made public before the first vote. --- (Right) Candidates ($N=28$) in the district of \Monthey{} and their relation to criteria.}
  \label{tab:criteria}
  \setlength\tabcolsep{3pt}
\begin{tabular}{llc|cc}
 \toprule
                  &           & Number of   &    Number of   \\
    Criteria      &  Category & Seats & Candidates & (\%) \\
  \midrule
  Region & Region 1 & $\geq5$ & 14 & (50.0\%) \\
  Region & Region 2 & $\geq4$ & 7 & (25.0\%) \\
  Region & Region 3 & $\geq3$ & 4 & (14.3\%) \\
  Region & Region 4 & $\geq2$ & 3 & (10.7\%)\\
  \midrule
  Gender & Male & $=8$ & 16 & (57.1\%)\\
  Gender & Female & $=9$  & 12 & (42.9\%)\\
  \midrule
  Age & 18-30 & $\geq4$ & 8 & (28.6\%)\\
  Age & 31-65 &  $\geq7$ & 16 & (57.1\%)\\
  Age & +65 & $\geq4$ & 4 & (14.3\%)\\
  \midrule
  Total seats & & $17$ & & \\
\bottomrule
\end{tabular}
\end{table}

\begin{figure}
    \centering
    \includegraphics[width=0.4\textwidth]{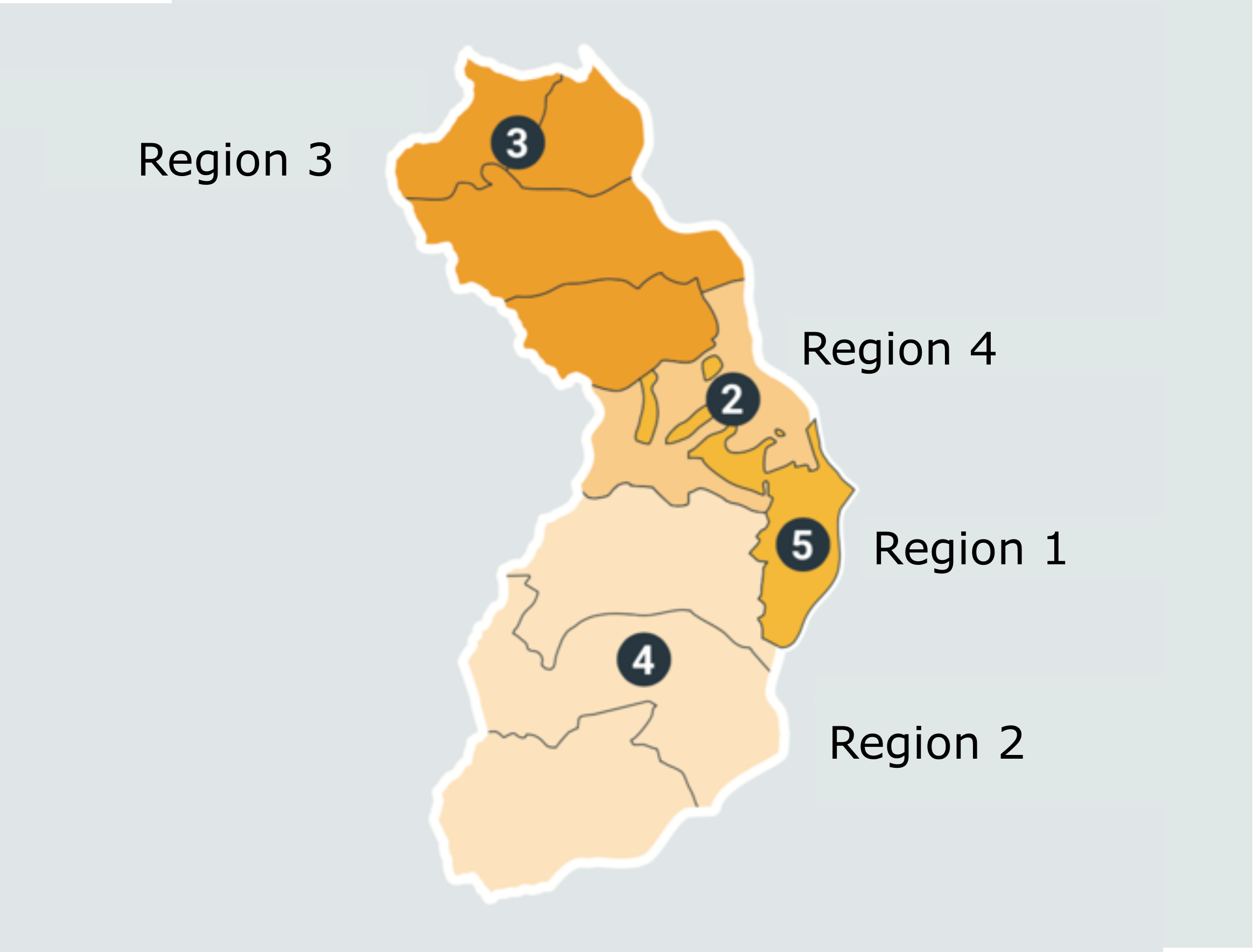}
    \caption{Use case --- Map illustrating the lower threshold of the region criterion for all regions in the electoral district of \Monthey{}. This information was made public before the first vote.}
    \label{fig:monthey}
\end{figure}

\subsection{Detailed Results in \Monthey{}}


The district of \Monthey{} had 28 candidates for 17 seats.
{
This number is already interesting. For many political organizations, it is difficult to convince people to be candidate on a list. As explained by candidates after the election, many decided to be candidate because of the Representation Pact mechanism. For instance, C.R. (a female candidate) explained in an op-ed of a national newspaper 
that she decided to be candidate ``because the mechanism guaranteed true opportunities for women to be elected on the candidates list. For once, we were certain not to be asked to be candidate exclusively to give an impression of equality'' \footnote{\url{https://www.lenouvelliste.ch/constituante/constituante-la-primaire-numerique-d-appel-citoyen-a-rendu-son-verdict-782906}}. 
}
A summary of the candidates relation to the criteria is depicted on the right-hand side of Table~\ref{tab:criteria}. 
The integer linear program corresponding to that use case is presented in Fig.~\ref{eq:use-case} in the Appendix. An interesting observation is that there were exactly as many candidates aged more than 65 as the number demanded by the criteria. In other words, in order to secure age diversity as foreseen by the criterion, these 4 individuals were automatically elected. Thus, it would leave only 13 seats open in the competition between candidates of the other age groups. 

Variants of this situation, where candidates were  ``protected'' by some representation criteria appeared in other districts: for the Age criterion it appeared in 4 districts overall, for the Gender criterion in 4 districts, and for the Region criterion in 3 districts. A total of 31 candidates in all 8 districts were protected by some representation criteria, and thus were certain to be elected. Out of a total of 96 seats available, 65 were thus contested (67.7\%) while 31 others were attributed ``automatically'' based on the criteria and available candidates. 

Following the plurality-at-large scheme, each of the 331 voters in the district of \Monthey{} could choose up to 17 candidates on their ballot (number of seats allocated to the district of \Monthey{} for the general election). A total of 1,931 votes were cast (average 5.8 per voter). Results of the votes, along with the values of the criteria for each candidate are shown in Table~\ref{tab:monthey-results}. The last column contains 'Yes' if the candidate was elected by applying the method described in Section~\ref{sec:pact}. The winning list is the one that, among all possible combinations of candidates who meet the criteria, has the most total votes received by the winning candidates. The method used (ILP) 
mathematical tools
to find the winning list. 

\begin{table}[t]
    \centering
    \caption{Use case --- Results of the vote in \Monthey{}, sorted by number of votes received. The two candidates who received the least votes (and were not elected) are not shown.}
    \label{tab:monthey-results}
    \setlength\tabcolsep{3pt}
    \begin{tabular}{lcccrc}
    \toprule
Candidate   & Gender & Age   & Region & Votes & Elected? \\
\midrule
Candidate A & M      & 31-65 & 1      & 166   & Yes      \\
Candidate B & F      & 31-65 & 1      & 128   & Yes      \\
Candidate C & F      & 31-65 & 2      & 121   & Yes      \\
Candidate D & M      & 18-30 & 1      & 114   & Yes      \\
Candidate E & F      & 31-65 & 1      & 111   & Yes      \\
Candidate F & F      & 18-30 & 3      & 92    & Yes      \\
Candidate G & F      & 31-65 & 2      & 90    & Yes      \\
Candidate H & M      & 31-65 & 4      & 89    & Yes      \\
Candidate I & M      & +65   & 4      & 75    & Yes      \\
Candidate J & F      & 31-65 & 1      & 73    & Yes      \\
Candidate K & F      & 18-30 & 2      & 73    & Yes      \\
Candidate L & F      & 18-30 & 2      & 70    & Yes      \\
Candidate M & M      & +65   & 1      & 70    & Yes      \\
Candidate N & M      & 31-65 & 1      & 64    &          \\
Candidate O & M      & 31-65 & 1      & 58    &          \\
Candidate P & F      & 18-30 & 1      & 57    &          \\
Candidate Q & F      & 31-65 & 2      & 56    &          \\
Candidate R & M      & 18-30 & 1      & 56    &          \\
Candidate S & M      & 31-65 & 3      & 49    & Yes      \\
Candidate T & M      & +65   & 1      & 47    & Yes      \\
Candidate U & M      & 31-65 & 1      & 45    &          \\
Candidate V & M      & 31-65 & 1      & 45    &          \\
Candidate W & F      & 31-65 & 3      & 45    & Yes      \\
Candidate X & M      & 18-30 & 3      & 42    &          \\
Candidate Y & F      & 18-30 & 2      & 29    &          \\
Candidate Z & M      & +65   & 1      & 27    & Yes     \\
\bottomrule
\end{tabular}
\end{table}

Even if there is a mathematical guarantee that the winning list output by the algorithm is the best list possible, it might be useful to look at the result in detail, in order to be able to explain in layman's terms the reason why certain candidates were not retained despite a higher number of votes. 
In our experience, candidates who were eliminated wanted to understand the reasons. This explainability also plays a role for the trust in the overall system. 
Intuitively, the results can be understood by reflecting on the criteria as a ``protection mechanism'' for the representation of certain categories
of the population. 
The fact that the top candidates in terms of votes are elected is not a surprise (candidates A to M). Things become more surprising starting with the first non-elected candidate, N. 
As discussed above, we knew in advance that the 4 candidates aged more than 65 (namely I, M, T and Z) had their election guaranteed by the criteria Age~$>65$. 
It turned out that I and M were in the top candidates, while T and Z were not; hence, we could say that T and Z were ``protected'' by their age. 
Table~\ref{tab:monthey-partial} presents a partial election result considering the 15 candidates whose election was ``explained'' so far: A-M who are top candidates, and T and Z who are protected by their age. 
As shown there, when we consider only those candidates, the criteria of Region and Gender are not met. To meet them, there should be two more people from Region 3 in the winning list, as well as 1 male and 1 female. 
Given this observation, it becomes clear why candidates S and W (the male and female candidates with the most votes from Region 3) are selected in the winning list at the expense of other candidates who received more votes. 
This kind of explanation, even if it cannot be used to derive a solution in all cases, is readily understandable and helps unlucky candidates better understand the reasons behind their defeat.

It is also interesting to analyze the impact of the criteria application in terms of the difference of votes between the winning list according to the Representation Pact and what would have been the winning list without it. 
The top 17 candidates if we consider only votes and not criteria obtain a total of 1,507 votes out of the 1,931 votes cast, which means that 424 votes ($1931-1507$) were ``lost'' (or 22\% of the total votes). 
The optimal winning list, considering criteria, receives 1,440 votes (491 ``lost'' votes, or 25.4\%). The difference is 67 votes (3.4\% of total votes) between the two situations. This difference can be thought of as the ``price'' of respecting the three criteria. 
In a more positive way, it can be described as an ``investment'' in meeting representation criteria. The commitments formulated by the voters in the first vote of the Pact (the three criteria) require an investment if they are to be met. In our use case, the investment was the largest in the case of \Monthey{}. In other districts, the difference was lower; in fact, in 4 districts, there was no difference at all between the optimal winning list and the top candidates without considering criteria.

This novel mode of election received a great deal of national media attention. The originality of the process and the use of a participatory computational system was particularly discussed by political observers and commentators \footnote{\url{https://www.rts.ch/info/regions/valais/9892696-lappel-citoyen-pour-la-constituante-valaisanne-nest-pas-contre-les-partis.html}}\footnote{\url{https://www.nzz.ch/schweiz/einmal-verfassung-bitte-ld.1432190?reduced=true&_x_tr_sl=auto&_x_tr_tl=en&_x_tr_hl=en-US&_x_tr_pto=wapp}}\footnote{\url{https://www.letemps.ch/suisse/valais-appel-citoyen-lance-primaire-constituante}}.


\begin{table}[]
\centering
    \caption{Use case --- Status of criteria considering the partial election result with candidates A-M, T and Z only. The region and gender criteria are not met. In order to meet them, 1 male and 1 female candidate from region 3 should be nominated.}
    \label{tab:monthey-partial}
    \centering
    \setlength\tabcolsep{6pt}
    \begin{tabular}{lccr}
     \toprule
     Criteria & Target & Value &  \\
     Category & Value & Reached & Difference \\
      \midrule
      Region 1 & $\geq5$ & 8 & $+3$ \\
      Region 2 & $\geq4$ & 4 & $0$ \\
      Region 3 & $\geq3$ & 1 & $\mathbf{-2}$ \\
      Region 4 & $\geq2$ & 2 & $0$\\
      \midrule
      Male   & $=8$ & 7 & $\mathbf{-1}$\\
      Female & $=9$ & 8 & $\mathbf{-1}$\\
      \midrule
      18-30 & $\geq4$ & 4 & $0$\\
      31-65 &  $\geq7$ & 7 & $0$\\
      +65   & $\geq4$ & 4 & $0$\\
    \bottomrule
    \end{tabular}
\end{table}

\section{Guidelines and Recommendations}
\label{sec:guidelines}

In order to make the Representation Pact more implementable, we provide the following guidelines and recommendations for each step in the process.

\subsection{Vote 1: Voting on Criteria}

The formulation of the potential criteria should itself be a product of public consultation. It is a crucial part of the objective to make the computational system participatory.  
The population might also be invited to directly propose criteria. 
One could imagine that the population proposes criteria through petitions signed by a minimum number of citizen, or that a  group of representative people is formed (e.g., through random draw) and asked to propose criteria. Alternatively, an existing group (e.g., from the parliament or the government) could propose a set of criteria for public consultation. 
What is crucial is that the pre-selected criteria be made explicit before the formal vote on accepting or rejecting them. 
This includes making sure that every potential candidate can be integrated in an exclusive category, and that candidates would be able and willing to provide the information required to categorize them. 
Second, it includes making the implications of the criteria explicit. 
These implications shall be formulated as clearly as possible, if possible with visual support. For instance:  ``Gender - the winning list in region X will be composed of 8 men and 8 women''.

\ccolor{
From a democratic point of view, the main challenge remains to define which criteria 
should be put to vote. 
%
This includes deciding which criteria are important (meaning which representation do we want to secure) and, procedurally, who should have the right to define which criteria are put to vote ~\citep[p.~151]{stojanovic_dialogue_2013}. 
We address these difficulties 
by proposing a participatory approach: where possible, the voters themselves should define the list of criteria put to vote in the first phase of the Representation Pact.
%
%
To be suitable, each criterion must be expressed in terms of specific and mutually exclusive categories. 
In other words, each candidate must be able to be uniquely characterized by one and only one of the possible categories of each criterion.  
%
Furthermore, for the Pact to be fulfilled, the target values for each criterion must be specified in a clear and precise manner. 
For example, if ``gender'' is accepted as a representation criterion, one needs to define if parity is the objective, or whether there will be an acceptable range (e.g., 40\%-60\%). 
%
%
}

\ccolor{
    For simplicity, our methodology for criteria selection in the case study provided the voters with options regarding representation of certain pre-selected attributes.
    However, there can be other different methods
    that provide the voters with more control over the criteria design.
    %
    For example, voters can also be involved in choosing the attributes to be used to in criteria (such as ethnicity, region, etc.) and also in the process of deciding the taxonomy of the attributes (such as relevant ethnic categories or regional divisions) through surveys.
    %
%
%
Once determined, the potential criteria should be the object of a public debate (either publicly, or within the organization). This means that the organizers of the vote should plan some time to allow people to debate and establish an opinion on the criteria they want to choose.
Optionally, when voting on the criteria, it might be useful that voters additionally rank the criteria in order of preference. Thus, if no winning list exists that meets all the criteria, it may be possible to relax either criterion based on this order of preference. Note that this ``flexibility'' procedure and its implications must also be communicated in advance. For example: ``if, in a given category, there are fewer candidates available to satisfy the criteria
for that category, we will redefine the criteria as the number of actual candidates in that category and declare the remaining seats as ``free seats'', or ``if the age criterion cannot be met with the list of candidates, 
it will be removed altogether, as it was the least preferred criterion''.
}


{
In our experience, limiting the number of criteria to three seems reasonable from a cognitive perspective; it would allow for simple understanding and verification of the solutions and, from a design viewpoint, it simplifies the process of preference elicitation from the voters. 
%
%
These numbers also allow for the algorithmic solution to be executed quickly (in a matter of seconds). In general, the more candidates, the more winning slots, and the more criteria, the more complex the solution is. In fact, the problem is
``NP-hard'', for which no efficient algorithms are known. However, if the number of criteria remains relatively small (e.g., 10 or fewer), the algorithmic solution mentioned above appears to give an answer in a timely manner (e.g., within an hour). 
Furthermore, the complexity of finding the solution does not depend on the number of voters, and hence the proposed framework can find efficient solutions for large voter bases with small number of criteria.
Importantly, the framework will always give an optimal solution, the bottleneck is the computation time.
%
%
It is crucial to ensure this ahead of time, as the election results must be reported in a timely manner. This is easy to verify a priori via simulations to ensure that no computational limitation will arise at the time of an election.
}

Additionally, more criteria makes it more difficult to find enough candidates to ensure that winning lists that satisfy the criteria are even possible. If there are not enough candidates, or the candidates are not sufficiently diverse, no outcome will be able to satisfy the desired criteria irrespective of the number of votes.
One solution is to recruit more diverse candidates in the next step; indeed, one can use the same algorithm to determine whether the candidate pool is sufficiently diverse. 
{However, we also recommended securing some flexibility on the criteria as part of the election. 
This can be achieved in various ways. 
Rather than fixing exact numbers, the criteria can be defined as a lower threshold and could take the following form (taking age as example): ``there will be at least 2 people aged 18 to 35, at least 2 people aged 65 and over, and at least 4 people aged 36 to 64 among the 10 elected.'' This leaves 2 ``free'' seats, and gives more flexibility to potential solutions; the 2 remaining places will be given to the candidates having the most votes, independently of their age category (assuming, of course, that the lower threshold for all categories is reached).
While the above process would lead to a relatively-flexible criteria, it can lead to increased survey length and complexity. 
Balancing survey complexity against drawing out sufficient details to construct flexible representation rules will be context-dependent but, nevertheless, solvable with a careful design that takes the population demographics into account.
}

\ccolor{
The same participatory approach should extend to ``tie-breaking rules''.
%
Participants must be actively consulted on a mechanism to be used in case of perfect equality between two winning committees to break ties. As foreseen by many electoral regulations, it can be decided to draw lots for the sake of choosing the winning committee. Alternatively, a second-round vote in which the voters should choose their favorite winning committee could also be organized. The key requirement is to prevent arbitrariness in determining the final winning committee.
}

{A criteria once selected 
can also be re-used in future elections if it is deemed to be appropriate and if the population demographics haven't changed significantly since the first vote on the criteria.
This would ensure additional stability in the composition of the chosen committees, while allowing the voting population to experience the impact of the criteria and build trust/suggest modifications in future applications of the criteria.
Legally securing representation rules for future elections is a common practice in many counties; as mentioned earlier, representation criteria are often encoded in the national Constitutions (for example, reservation provisions in India \cite{jensenius2013power} and preselection quotas in Brazil \cite{jones2009gender}).
%
}


\subsection{Between Votes: Getting Candidates}

By the time the criteria are decided, it is not necessary to know who the candidates are or will be. Indeed, this might impact the criteria chosen.  The first choice should be made \textit{in abstract}, keeping in mind the representation which the majority wants to achieve, and not the election of a particular person. 
{
Once these criteria are defined, the organizers of the vote should make sure that there are enough candidates. The challenge is directly linked to the criteria chosen. The number and diversity of candidates should make a real choice possible. If a criterion can only be achieved with a single possible winning list, the choice is not really a choice. The organizers of the vote should stimulate and look for candidates with the general objective of bringing diversity among the pool of candidates. 
Our experience shows that this way of stimulating candidates offers new chances to attract people usually less interested in being candidates (e.g. women, young people, members of minority groups). The clear and explicit commitment taken on representation criteria shows that the relevant voters want diversity and that their candidatures are welcome, indeed necessary, to achieve this commonly shared goal. In our experience, this has an important positive impact on motivation to run as a candidate. As shown by the testimonial provided previously, candidates of underrepresented groups enjoy this clear commitment and associate it with fair(er) chance to get elected. 
}

{
Indeed, similar positive strategic behaviour has been observed in prior economics literature.
Inclusion of diversity rules has been shown to incentivize additional investment of effort to achieve the desired diversity \cite{fryer2013valuing,coate1993will,hu2018short}. 
%
However, negative strategic behaviour, in the form of attempts to \textit{manipulate/control} the results of a multi-winner election can also occur; such behaviour has been studied in the context of other proportional representation or tie-breaking rules \cite{obraztsova2013manipulation,procaccia2007multi,meir2008complexity}.
%
%
%
%
To prevent manipulation of criteria or final selection (e.g., forcing a selection of criteria that guarantees election of certain candidates),  as discussed in the previous paragraph, the organizers should encourage and reach out to diverse candidates with the goal of ensuring representation and multiplicity in the candidate pool as well.
%
}


Before the election, the organizers of the vote should publish the list of candidates with their categorization according to the various criteria. This information represents the basis of the vote to come. It must be transparent and understandable to voters. In addition, publishing the list allows anyone to verify the classification of candidates so that any potential mistakes can be identified and corrected before the election.

\subsection{Vote 2: Voting on candidates}
{
In addition to all the conventional requirements of a fair election, the voting of candidates should fulfill further requirements. 
First, the election need not be digital, i.e., it can employ the same analog/digital tools used for other elections in the region. }
What is crucial is the capacity to digitize the raw results (the individual votes which every candidate has gotten). These raw results would then be used by the algorithmic counting method to find the winning list. 
Second, this raw results list should be published in an open format (e.g. CSV). If possible (e.g., if the election was done using an e-voting platform that allows it), the raw individual ballots themselves could be published as well in an open format. 
In some e-voting systems, voters receive a unique code or 'hash' after their vote, which is associated with their individual ballot and allows them to check from the full ballots list that their vote was taken into account. 
In analog elections, people should be able to attend the counting as observers.

{
Along with the above steps, many other actions can also be undertaken to ensure that the election process is open and transparent.
For instance, the source code of the algorithm used to identify the winning list should also be accessible to the public. In combining the raw results list with the algorithm, it should be possible for anyone to verify the outcome of the election 
or observe counterfactuals.
In fact, independent third party organizations, who have sufficient background with the tools used for the election process (i.e., ILP solvers), should be appointed to audit the election results.
}
{To further improve the transparency of the process, open-source solvers should ideally be employed to solve the computational problem.
However, ILP solvers currently used in practice are often proprietary and the mechanisms by which the results are obtained are not always verifiable by the general population.
In light of this shortcoming, 
%
auditability can be established 
by verifying the results using multiple ILP sovlers from different sources.
%
Trust in this process is greatly enhanced by taking actions to ensure transparency and can mitigate the challenges faced in communicating the results.
}

\section{Discussion and Limitations}
\label{sec:discussion}

The Representation Pact supports elections that respects binding, modifiable and legitimate criteria.
In this section, we discuss the criticisms and possible extensions of the Representation Pact. The criticisms are interesting lessons learnt for real-world applications of computational systems in politics, specifically in electoral processes.


\subsection{Political Criticism of the Representation Pact}

One key argument against the Representation Pact is the idea that candidates with more votes should always be selected.  
Alternatively expressed, it is criticized that an ``old'' candidate with fewer votes gets elected only because of his age. There is a classical issue linked to representation, and a newer issue linked to the use of a computational system. 
On the first point, it is important to underline that the Representation Pact does not create something completely new. It rather amplifies a logic already present in the current system and makes it more dynamic.
In a proportional election, it is often the case that an elected representative of a certain party has obtained fewer individual votes than a non-elected candidate from another party. The same is true for distinct electoral districts in which candidates of smaller districts get elected despite having secured far less votes than candidates from a more populated district. 
These situations do not come out of the blue. They are based upon the choice made by citizens and, as such, can be reconstructed along the two democratic moments of the election: the choice of criteria (political or geographical representation) and the election itself. These two moments must be thought of as a whole.
However, the criticism can be expressed in the form of an ``essentialist'' criticism, as explained above. An ``old'' candidate gets elected only because of his age. In that sense, the candidate is reduced to one feature of their overall identity~\citep{mansbridge_quota_2006}. The dynamism --- understood as the capacity by citizens to change the chosen criteria and their implications --- is a key argument against the dangers of essentialism~\citep{mansbridge_quota_2006}. Individuals, in our case candidates for an election, are not permanently defined by and through specific socio-demographic features. Criteria are chosen for a specific election and might  change for the next election or for an election in another institution.  
On the second point, it is interesting to highlight that the well-known tension around representation 
seems to be reinforced by the use of a computational system. One local political commentator reacted through an op-ed and claimed that "candidates were chosen by a robot". This challenge relates to scholarship focused on the explainability and transparency challenges of AI applications~\citep{lee_understanding_2018}. Above all, this use-case reinforced the lesson that the algorithmic output should be transparent and explainable to the candidates
and to the broader public.

{
Explainability and transparency can have several objects. First of all, it is relatively simple to make sure that one's vote has been duly taken into account. In that sense, transparency can be achieved by publishing anonymized ballots and providing access to source code of the system.
This might be especially important in the context of an e-election, 
although our approach does not require e-voting. Secondly, explainability and transparency can bear upon the result (the winning committee). It is important to underline that the fact that the winning committee fulfils the prescribed criteria can easily be explained. In that sense, the public can verify if the promise made during the first phase (definition of criteria) is fulfilled. The true explainability challenge bears upon the mathematical process identifying the winning committee.  Explainability is here not a challenge similar to complex machine learning settings that resist explanation; it is here mainly about communicating in layman's terms the mathematical work done by the computational tool. In that sense, the challenge is similar to other electoral processes relying on complex mathematical operations (such as bi-proportional election mechanism used in \Valais{} or counting processes for ranked voting methods). While the underlying principle of the counting methods are easy to understand, the actual implementation might be difficult and the counting might be practically impossible without the help of a computer. 
The experiences from 2017 French presidential election about alternative electoral mechanisms (e.g. ranked voting with instant-runoff counting rule) can be considered as example of this communication challenge. 

We identified three lessons learnt from our case-study. Firstly, there is communication challenge to explain that  the fulfilment of the criteria is aimed at the list's level, and not at the individual level. Secondly, the design and presentation of results can be instrumental in providing clues explaining the winning of certain candidates. We invested time and resources in designing the presentation of the list of results in a easily understandable way. Furthermore, we also carefully explained the general mathematical work done by the counting process with simple examples, and we made this communication material broadly available. Thirdly, it would be good to develop an add-on allowing individuals to play with hypothetical scenario and test how their own socio-demographic characteristics with respect to the criteria affect their chances to get elected.
We have created such an add-on allowing to test different scenarios with respect to a past national election. These pragmatic measures can address the communication challenge. 
However, 
there is still an important issue of trust at stake: candidates, voters, but also the broader public need to trust the mathematical method that identifies the winning committee. To foster this trust, it is important to explain that the mathematical method has been the object of scientific, peer-reviewed publications and has been, as such, verified by independent experts who do not have any stake in a particular election. Moreover, publishing anonymized raw ballots and the parameterization details of the ILP allows anyone competent to verify the results using the solvers of their choice. The transparency of the voting data and the reproducible nature of the results are key in building trust around the whole election. 
}

{
%
%
%
%
}

\subsection{A Social Contract for Elections}

The two worries addressed in the precedent section calls for a more affirmed political commitment to the values underlying the Representation Pact. This is what we call a social contract for elections, taking into account the specific challenges linked to the use of a computational system.  We present here preliminary content for this social contract. 
It is important to underline that the two moments of the Pact refer to different sets of values.
The first vote on the criteria concerns mainly the socio-demographic characteristics of the candidates. It refers to the values of descriptive representation and diversity as previously explained. We need to answer the following questions: (i)
according to which aspects of identity should elected officials be representative of the population? (descriptive representation), and (ii)
according to which aspects of identity should the people elected be as diverse as possible? (diversity)
.


In contrast, vote 2 refers mainly to the political positions of the candidates. When it comes to giving their votes, voters choose according to their political preferences. Candidates are judged for the projects and ideas they defend. 

The Representation Pact helps to bring order to the way these criteria are usually addressed as part of political debates.  Today, in an election, political preferences are often draped in the language of  ``competences''. One then juxtaposes  ``quota'' with  ``political competence''. Gender parity pays a particularly heavy price for this mix of categories. Representation from a gender perspective is systematically opposed to competence. Strangely, we hear less of this discourse on geographical criteria, where elected representatives from one district or another would not be chosen for their competences, but merely upon the location where they happen to reside. 
The Pact shows that this way of handling these criteria is flawed because it mixes two different levels. The representation criteria (``quota'') relate to socio-demographic factors and are the object of vote. Political positions - a more honest term than  ``political competences'' which suggests that this is purely a resume-like issue - are at the heart of vote. 
This way to frame the two levels as the dual components of a complete democratic moment is an interesting reply to the technology-fueled worries described above. This framing makes clear that the first vote is a choice about representation, without any impact by technology. The second choice is also a classical election moment. Only the linking of the two votes as two components of a complete democratic choice requires a computational system.

\subsection{Possible Extensions}

The idea behind the Representation Pact can be extended to other situations. 
The Pact may also apply to companies or associations. When choosing a Board of Directors or committee, shareholders or members can first vote on the criteria and then proceed to the election. It is interesting to recall that the representation criteria could indeed change according to the specific body at stake. Thus, the importance of representation according to certain criteria (gender, age, ethnicity, income or other) might be considered more important for a legislative body (a parliament) than for an executive. In the same way, the Board of Directors in a business company might not be assessed according to the same criteria as the committee for a local NGO. The strength of the first moment of choice of criteria lies in its capacity to adapt to specific social realities in which representation might be differently defined.

%
{
The use of representation criteria can also be extended to other voting systems.
For example, diversity constraints have been proposed for approval voting \cite{brams1990constrained} and proportional representation \cite{lang2018multi}.
However, employing them for complex voting systems, like preferential voting, can be more difficult, since the objective function may not simply count the number of votes received by each candidate, but also the preference rank assigned to each candidate by a voter.
}
\section{Conclusion}
We have outlined the theoretical background and a case-study of the computational system we called the Representation Pact. Citizens agree collectively on a representation criteria and apply them to elections. 
Expectations in terms of representation are an important challenge for democratic decision-making bodies. In providing a democratically legitimated solution, this approach allows democratic bodies to become profoundly diverse and dynamic. It contributes to fair political elections in securing descriptive representation and diversity in specific circumstances. The approach has proven successful in a real-world example of a political party's primaries in \Switzerland{}. 

{
The Representation Pact is grounded in participatory design principles; however, it is important to note that the presented design in the use-case is one way to actively engage voters in a diverse committee selection process.
%
Voting and elections are complex processes and, as mentioned earlier, highly context-specific. A two-phase voting process can be feasible in certain settings and can be resource-constraining in others.
Our goal of presenting the Representation Pact is to propose a complete framework that returns a diverse and popular committee; context-specific modifications to this framework should be pursued as long as they satisfy the inherent philosophical and participatory principles that motivate our framework.
}
The fact that the Pact is generic promises a high potential impact, as it can be applied in many different election contexts, including political party primaries as described in the paper, but also in the election of non-political bodies (boards in associations, academic institutions, corporations etc.). \ccolor{Finally, the lessons learnt from this experience can inform the implementation of other computational systems in the realms of politics.}


\clearpage
\bibliographystyle{ACM-Reference-Format}
\bibliography{cscw-2021}
\clearpage


\appendix

\section{Details of the Integer Linear Program} \label{sec:ilp}

The method employed to count the votes and find the winners is based on previous research~\citep{ijcai2018-20} in which randomized approximate  solutions were developed for this problem under a wide class of scoring rules.
In essence, the paper tackled more complex ways to aggregate votes beyond the plurality-at-large method employed in \Switzerland{} and in many countries around the world.
Although its solution is general and widely applicable, it poses two problems in this context.
One is that the approach it suggests is often a randomized solution; the same results of vote 1 and vote 2 could end up with different winning committees depending on the outcome of a few coin flips.
This would be understandably unsatisfactory for the candidates and public at large.
Furthermore, the approach it suggests is approximate; in other words, it may not result in a set of winners that maximizes the number of votes collected, rather it just guarantees that the number of votes collected by the winners is close to the maximum.
This is again unsatisfactory, in particular for a candidate that would have been in the optimal winning list but is not present in the approximate list.

To combat this, we derived a special case for the approach which can be solved deterministically and exactly for the parameter ranges in question.
%
Formally, we have $m$ candidates $C=\left\{1,\ldots,m\right\}$, a total of $n$ voters and a desired size $k\in \left\{1,\ldots,m\right\}$ of a committee to be elected. 
For the purposes of illustration, let's say each voter can select at most $k$ candidates, though this restriction can be changed without consequence. 
Let's say candidate $i$ receives a total number of votes $w_i$.
A committee $S\subseteq C$ of size $k$ is considered to be good if it receives many votes, i.e., if the sum of all of the votes for the candidates in S, $\sum_{i\in S} w_i$, is large. 
Overall, our goal is to elect a committee $S\subseteq C$ of size $k$ that satisfies the criteria selected. 
Furthermore, amongst all possible committees that do so, we would like to elect the one with the most votes, i.e., with maximal $\sum_{i\in S} w_i$.
We can encode this problem mathematically by using a $0/1$ variable $x_i$ to represent whether the candidate $i$ is selected ($x_i=1$) or not $(x_i=0)$. 
If candidate $i$ is selected, then it contributes $w_i = w_i\cdot x_i$ votes to the committee (as $x_i=1$).
Otherwise, it contributes $0=w_i\cdot x_i$ votes (as $x_i=0$).
Thus, the total votes received by the selected committee is captured by the equation 
$\sum_{i\in C} w_i\cdot x_i.$
%
Using this formulation, we can write down an integer linear program (ILP) for a particular use case to encode the mathematical problem of finding the best committee that satisfies this criteria.
For example, in the use case of Table~\ref{tab:criteria}, the ILP would be as shown in Fig.~\ref{eq:use-case}.
%
%

\begin{figure*}
\begin{centering}
\begin{align*}
\label{eq:ILP}
\text{maximize}~& \sum_{i\in C} w_i\cdot x_i & (\text{maximize the total votes})\\ 
\text{subject to}~& \quad  x_i\in \left\{0,1\right\} \quad \mbox{for all $i\in C$}, & (\text{$0/1$ variables})&\\
& \quad  \sum_{i\in C} x_i = 17, & (\text{committee size constraint})\\
& \quad \sum_{i\in P_{R1}} x_i \geq 5, \sum_{i\in P_{R2}} x_i \geq 4, \sum_{i\in P_{R3}} x_i \geq 3, \sum_{i\in P_{R4}} x_i \geq 2. & (\text{region constraints}) \\
& \quad \sum_{i\in P_m} x_i = 8, \sum_{i\in P_w} x_i = 9, & (\text{gender constraints}) \\
& \quad \sum_{i\in P_{18-30}} x_i \geq 4, \sum_{i\in P_{31-65}} x_i \geq 7, \sum_{i\in P_{+65}} x_i \geq 4 & (\text{age constraints}) 
\end{align*}
\end{centering}
\caption{Integer linear program (ILP) for the use case presented in Table~\ref{tab:criteria}. 
{Any choice $\{x_i\}_{i \in C}$ represents a particular committee (consisting of candidates for which $x_i=1$) and, correspondingly, the objective function $\sum_{i\in C} w_i\cdot x_i$ captures the total votes received by this committee. 
The inequalities capture the constraints to be satisfied by the chosen committee. 
The first two constraints are structural, limiting $x_i$'s to take only 0 or 1 value, and restricting the size of the committee (17 in this case).
The last three constraints characterize the chosen representation criteria in Table~\ref{tab:criteria} (with respect to region, gender, and age attributes).
The committee that satisfies all the above constraints and achieves the maximum number of votes (amongst all \textit{feasible} committees) is the solution to this ILP program for this setting.
}
}
\label{eq:use-case}
\end{figure*}

%
%
%
%
In general, solving an integer linear program is NP-hard and hence can take a prohibitively long time. 
{However, the (in)feasibility of executing the above program in real-world settings is dependent on a variety of factors and, in many cases, the optimal solution can be found in a small amount of time.
A straight-forward way of solving the ILP is to simply check the feasibility and compute the number of votes assigned to all possible $2^m$ committees.
The processing time required for each iteration in this algorithm is polynomial in number of constraints in the chosen criteria and the number of voters $n$.
Therefore, the main bottleneck in the complexity of solving this program is indeed the number of candidates, which in most elections is not large and can be handled using standard ILP solvers.
%
Since size of voter-base is not the determining factor, the standard ILP solvers should be able to handle settings with large voter-bases as well.
Beyond the above described method of solving an ILP, there are also well-studied approaches in optimization literature that are significantly faster than the trivial approach.
Indeed, 
%
we use the well-known and publicly available packages, such as \textbf{CPLEX} or \textbf{CVXPY}, which use heuristic methods (such as, ``branch-and-cut'' approach) to speed up the computation while still ensuring that we get the optimal solution -- this allows us to produce a result within seconds, even when there are a large number of candidates.
The time taken to generate the solutions for the use case described in Section~\ref{sec:usecase} is less than one second.
%
We also provide an example implementation that is publicly available here: \ANONURL{}.
}

\section{Philosophical and Participatory Motivations for Representation Pact} \label{sec:motivation}
{The Representation Pact aims to ensure that a democratically-elected committee is diverse and representative of the underlying population.
In this section, we first discuss the philosophical motivation behind the goal of achieving representation through the Pact.
Secondly, to realize this goal, it is important that the Pact is executed in an effective and thoughtful manner; to that end, we highlight the participatory  principles behind the design of the Representation Pact and show that following these principles can lead to successful implementations of the Pact.
}

%
\subsection{Philosophical Grounding}
The Representation Pact should be seen as an example of institutional mechanisms securing a specific representative outcome~\citep{mansbridge_should_2015}. We need to address the theoretical underpinnings of representation in order to highlight the normative goal which the computational system integrated into the electoral process is pursuing.
In a first political sense, representation is the attribute of someone who has been elected to represent a specific citizenry and who has obtained specific legislative or executive competences. This is the sense used in "representative democracy".
In a second socio-demographic sense, representation is about the characteristics of the people elected. In other words, we can raise the question as to whether elected representatives are similar (in significant ways) to the people who have elected them. This sense of 'representation' can be referred to as the 'mirror-' or 'descriptive representation'.
The main issue we raise is to ask in which situations, if any, this descriptive representation is a normative goal. To answer this question, we rely upon the conceptual framework proposed by ~\citet{mansbridge_should_2015}. As we shall discuss later, Mansbridge's approach has been the object of criticisms.  
To start with, it seems legitimate to assume that elected representatives should have the capacity to take into account the perspectives of constituents who do not display socio-demographic features similar to their own (such as gender, age, or place of residence). For instance, we do not assume that White representatives could not represent Black people's interests. Representatives could indeed take into account different people' interests. We should ask a more limited question: in which circumstances is descriptive representation a normative goal capable of addressing problematic situations \textit{despite} the general capacity of elected representatives to account for the perspectives of differing socio-demographic groups? According to~\citet{mansbridge_quota_2006}, there are two situations in which descriptive representation is important. 
 The first situation is when the interests and perspectives of a specific group cannot adequately be represented by elected representatives who are not members of that specific group. 
As explained by~\citet{mansbridge_what_2000}, this argument is especially important in constellations in which information or trust is absent or lacking. This is to say, constellations in which a) the interests of the group are not even crystallized as political issues,
b) past injustices and mistrust have made communication between members and non-members difficult, and c) the mere physical presence of members of the group fosters attention on the side of non-members and helps give members' perspectives a more important place.
In all of these situations, descriptive representation relies upon an instrumental argument: increasing the presence of members in the elected body increases the chance that these issues are dealt with. It hence fosters the quality of political work. As explained by Diamond and Morlino, the quality of political work in a democracy can be measured by the capacity of institutions to produce public policy which corresponds to the demands of citizens~\citep[p.~4]{diamond_quality_2004}. Assuming this broad approach, the situations described above are interesting because they add relevant voices to representative bodies and they add substantial arguments to political debates~\citep{bratton_effect_2002}.
The second situation identified by~\citet{mansbridge_quota_2006} is about using descriptive representation as an instrument to address  discrimination. In many constellations, discrimination patterns prevent members of groups from being elected, or even from being candidates. As proposed by Mansbridge, we could distinguish between surface and structural discrimination. Surface discrimination is relevant when it comes to electoral habits by citizens and the difficulty for members of specific groups to get elected because of negative \textit{a-priori} assumptions linked to socio-demographic features displayed by a group's members. This surface discrimination might even be unconscious, e.g., implicitly considering that women are less qualified for political functions. The structural discrimination is about conditions impacting the chances of getting elected. For instance, women are assumed to be the primary care-givers, leaving them with less time for political activities. In these constellations, descriptive representation is a powerful tool for guaranteeing that specific groups are represented, thereby changing individual and collective discriminatory patterns. The argument can indeed be reinforced by a justice-in-time argument, for instance if a group is owed a specific reparation for reasons of justice.
In the meantime, the ambition of descriptive representation has been the object of criticisms~\citep{meier_dark_2018}. Firstly, there is no guarantee that the elected person will represent the group's interests~\citep[p.~354]{young_justice_1990}. Indeed, the argument proposed here as an important empirical dimension in claiming that chances are better to see specific interests be defended. Secondly, there is a danger of essentialism in assuming that a group's members feature specific characteristics which are unreachable for non-group members. In an essentialist perspective, women's interests could only be defended by women because of the essence of being a woman. The reply to this line of criticism is to underline that the argument is focused on the experiences made by group's members, not on a putative biological or genetic identity. Likewise, this criticism underlines the danger of considering all group members as having similar interests. Mirror representation should not be used to minimize the intra-group diversity and the multiple, and mutually reinforcing patterns of discrimination found within a group should be taken into account (intersectionality).
In brief, this section has highlighted the theoretical context of the Representation Pact. This context provides the broader theoretical framework justifying the normative goal which the computational system is pursuing. 



\subsection{Participatory Design Grounding}
{
Traditionally the design of any technological system is governed by centralized organizations, keeping the end-users of the system in mind.
However, an isolated process, where the design decisions are entirely made by the governing organizations, often does not completely satisfy the needs of the end-users \cite{robertson2012participatory}.
Participatory 
design techniques 
address this problem by actively involving end-users in the design process so that the final \textit{end-product} focuses on the needs and requirements of the users \cite{schuler1993participatory, spinuzzi2005methodology}.
In general applications, such design methods are expected to satisfy the principles of including diverse stakeholder opinions in the design process and ensuring a balanced power dynamic between users and designers of the applications; we argue that a proper implementation of Representation Pact indeed follows these principles as well.
%

\noindent \textbf{Active involvement of stakeholders in the design process.}
{
    Elections are already a way for voters to exercise their option of actively participating in a democracy and the final chosen committee composition.
    However, by providing a democratic mechanism to choose the criteria as well (via the first phase of Representation Pact), the voters are provided with power to not only choose the \textit{winners} of the elections, but also the power to determine what the chosen committee as a whole should look like.
    %
    %
    %
    Nevertheless, 
    simply involving voters in the criteria-selection process is just one step towards building trust in the democratic process for all parties involved.
    %
    %
    Transparent channels of communications between the voters and the institutions governing the elections (as discussed in Section~\ref{sec:discussion}) is also necessary for a successful implementation of the Representation Pact.
}

\noindent \textbf{Ensuring appropriate representation.} 
{
    Participatory design techniques, by intention and impact, lead to better representation of various different groups of the population in the design process
    \cite{spinuzzi2005methodology}.
    %
    The intention and impact of Representation Pact is similar. The first phase ensures that the intention to choose a diverse committee is incorporated by asking the voters to participate in design of the selection criteria.
    The final phase finds the committee that has the most votes, while satisfying the criteria determined during the first phase.
    %
    %
    Given a robust and voter-representative committee criteria, the impact of this selection process is that the chosen committee appropriately represents the voting population as determined by the committee criteria.
    The importance of representation in committee selection cannot be understated.
    Significant amount of research in stereotype propagation and cultivation theory has shown that the lack of appropriate representation of marginalized groups in selection processes can induce new or propagate existing biases against these groups \cite{spencer1999stereotype, word1974nonverbal, shrum1995assessing, gerbner1986living, kay2015unequal, collins2002black, harris1982mammies}.
    Ensuring appropriate representation of marginalized groups in democratically-chosen committees is a step towards addressing the broader discrimination faced by such groups.
    }

    \noindent \textbf{Simplicity and transparency.} 
    {
    Several applications and case-studies on co-operative design processes have highlighted the importance of simple and effective communication with the users \cite{mayer2013lessons}.
    Often the challenges in the application of such design methods
    arise from the difficulty of communicating the technical aspects of the product to the end-users; without such communication, eliciting uncomplicated and useful feedback can be hard.
    %
    }

    {    
    Application of Representation Pact can face similar challenges as well.
    The first phase of preference elicitation from the voters on the desired group-representations in the committee can be tough to implement in an uncomplicated manner in every setting.
    As we discuss in Section~\ref{sec:pact}, the process used in our use case for voting on criteria presents the users with a series of yes/no questions around the group-representations in the committee and uses the aggregated responses to choose a criteria.
    The reason for presenting this voting process as a series of questions, each addressing representation with respect certain group memberships, is to ensure that the process of criteria selection is uncomplicated for the voters.
    Nevertheless, as mentioned earlier, design frameworks that involve the voters in the process of choosing the attributes used in the criteria can also alternately be employed. 
    Our goal in the construction of the first phase discussed in Section~\ref{sec:pact} was to reduce the underlying complexity of criteria selection, and other frameworks that satisfy this goal can also alternately be employed in the Representation Pact.
}    
    
    
\noindent \textbf{Culturally-relevant design.}
    While the above discussion highlights the use-cases of participatory design, the application of such design in governance is extremely dependent on the context \cite{merritt2012cultural}.
    For instance, \citet{hussain2012participatory} highlight the challenges of implementing participatory design techniques in Cambodia and discuss the differences in applications of such techniques between developed and developing countries.
    In the setting of elections, the challenges of implementation across different countries and culture will be further pronounced; correspondingly, the first phase of our Pact, i.e., voting to determine the criteria of representation goals is be really important towards ensuring that the final selected committee represents the voter-base.
    %
    %

The above discussion aims to highlight the participatory design principles followed by Representation Pact. 
Application of this pact (as discussed in Section~\ref{sec:usecase}) can lead to increased voter participation in the democratic process and ensure that the chosen committee represents the voter population in an equitable manner.
}

\end{document}